\documentclass[11pt,a4paper]{article}
\usepackage{jcappub}

\title{Dense DM clumps seeded by cosmic string loops and DM annihilation}

\author[a,b]{V.S. Berezinsky,}
\author[c]{V.I. Dokuchaev}
\author[c]{Yu.N. Eroshenko}

\affiliation[a]{INFN, Laboratori Nazionali del Gran Sasso, I-67010 Assergi (AQ), Italy}
\affiliation[b]{Center for Astroparticle Physics at LNGS (CFA),
I-67010 Assergi (AQ), Italy}
\affiliation[c]{Institute for Nuclear Research of the Russian Academy of
Sciences, Moscow, Russia}
\emailAdd{berezinsky@lngs.infn.it}
\emailAdd{dokuchaev@inr.ac.ru}
\emailAdd{eroshenko@inr.ac.ru}

\abstract{We develop a model of production of the very dense clumps of DM in RD epoch
due to the accretion of DM on the loops of cosmic strings as the seeds.
At some time the loops disappear, for example due to the gravitational
radiation, and the remaining dense clumps produce the enhancement of the 
annihilation signal. We take into account the velocity distribution
of the strings, and consider the two extreme regimes of DM annihilation: fast
decay and continuous evaporation. The produced annihilation
flux of gamma radiation is detectable, and for some parameters of DM
particles and the strings can exceed the extragalactic flux of the 
gamma-radiation observed by Fermi. For the fixed parameters of DM
particles (e.g. neutralino with fixed masses and cross-section of annihilation)
one can obtain the limits on the basic string parameter, tension $\mu$,
which is stronger than  (more general) limits obtained from WMAP
observations, cosmological nucleosynthesis and gravitational lensing. In
particular for the neutralino with 100 GeV mass we exclude the interval
$5\times10^{-10}<G\mu/c^2<5.1\times10^{-9}$.}

\toccontinuoustrue

\begin{document}

\maketitle
\flushbottom

\section{Introduction}

The linear topological defects --- cosmic strings can be formed in the early
cosmological phase transitions (see for a review \cite{VilSheBbook},
\cite{Vil05}). Along with the infinite strings there is possibility of
closed loops formation in the network of curved cosmic strings due
to their interconnections. According to numerical simulations,
after a long transient stage a true scaling regime sustained, then
the typical distances between the strings and the coherence length
both scale in proportion with the horizon scale \cite{alpha}. A
string loop formed at the cosmological time $t_i$ has the
length $l\simeq\alpha ct_i$, where in the scaling regime
$\alpha\simeq0.1$ according to \cite{alpha} and \cite{alpha2}, 
although other values of $\alpha$ were obtained in other works 
(see for example \cite{alpha3}, \cite{alpha4}), down to $\alpha\sim10^{-3}$.

A fundamental characteristic of the string is the mass per unit length
$\mu\equiv M_l/l$ or the tension, which is of the order of symmetry breaking
energy squared $\eta^2$. For example, the grand-unification-scale
strings have $G\mu/c^2\sim10^{-6}$, where $G$ is the Newton's
constant. There are several restrictions on $\mu$. From CMB
observations it follows $G\mu/c^2\le2\times10^{-7}$
\cite{PogWasWym06}. The bound $G\mu/c^2\le10^{-7}$ was obtained
from the study of nucleosynthesis \cite{alpha}. Search for the pairs of galaxies images consistent with
the gravitational lensing of the cosmic string presents the limit
$G\mu/c^2<3\times10^{-7}$ at 95\% confidence level \cite{Christiansen08}.
In \cite{OluVil06} the formation of stars in the first dark matter DM haloes seeded by the loops was considered. It was found that $G\mu/c^2<3\times10^{-8}$ to avoid collision
with WMAP data on the reionization redshift. Searches for the gravitational
wave bursts from strings by LIGO provide the joint constraints on
the strings parameters ($\mu$ and interconnection probability)
\cite{LIGO}, but $G\mu/c^2$ is only weakly restricted in comparison with
the constraints above. And finally, the strongest bound $G\mu/c^2\le4\times10^{-9}$ was obtained
from the pulsar timing \cite{RvanHaas11}.

In this work we present new constraint on $\mu$ which was obtained
from the DM particles annihilation in the dense clumps
seeded by the loops at the cosmological stage of radiation
dominance. The direct detections of DM particles are the promising but
still elusive experimental problems, therefore the search of
indirect signature of the DM is important for clarifying the DM
origin. The promising indirect signature --- DM particles annihilation would proceed
more efficiently (it can be boosted by several orders) if the
Galactic halo is filled by the dense DM substructures or DM
clumps. The cosmic string loops produce very dense clumps due to their early
formation. Only low velocity loops can produce the clumps. The
probability of the low velocity loop formation is very small, but
even tiny fraction of the formed loops may produce the dense
clump population and significant annihilation signal.

The clumps formation at the radiation dominated (RD) stage was studied in details by \cite{KolTka94}. In the
particular case of loops' density perturbations the maximum density
of clump is  restricted due to adiabatic expansion of the
already formed clump after the loop gravitational evaporation. We
found the modification of this restriction in the case then the loop decays before
the clump virialization. In this case the clumps can reach density
$\rho_{\rm cl}\gg140\rho_{\rm eq}$, where
$\rho_{\rm eq}$ is the density at equality. The comparison of the
resulting annihilation signal with the Fermi-LAT data allows us to obtain
the restriction on the string parameter $\mu$. It must be pointed
out that our constraints were obtained by supposing that the DM
can annihilate, and we take the $\sim100$~GeV neutralino as the
most promising particle candidate. The constraints will be
different for other DM models. In this sense the obtained
constraints must be considered as joint constraints on the
properties of the string loops and DM particles.

\section{Initial speed of the loops and rocket effect}
\label{velsec}

Here we consider the influence of the initial velocities of the loops and
rocket effect on the evolution of perturbations and clumps formation.

The necessary condition for the clump formation is the low
velocity of the loop \cite{KolTka94}.
Only results for average initial velocity of formed loops were
presented in literature but the distribution over velocities is
possible. We interested in low velocity end of the distribution,
because the clump forms only if the seed loop stays near the
center of the clump during its evolution. Loops can be formed by
intersecting long string segments or by self-intersection of long
strings. We suppose that the probability of velocity components of
the loops is simply Gaussian with mean value at the correlation
length scale $\langle v_i^2\rangle^{1/2}\simeq0.15c$ \cite{AllShe90}, and
therefore the probability of full initial velocity is
\begin{equation}
P(v_i)dv_i\simeq\frac{2^{1/2}dv_iv_i^2}{\pi^{1/2}\langle
v_i^2\rangle^{3/2}}e^{-v_i^2/2\langle v_i^2\rangle}.\label{distrv}
\end{equation}
Even if the process of a loop's formation involves the intersection of two strings
with large velocities, it does not necessarily means that the
resulting velocity will be high.

The displacement of the loop beginning from its birth moment $t_i$ till the
full decay moment $t_d$ is
\begin{equation}
\Delta
r=a(t_d)\int\limits_{t_i}^{t_d}\frac{v(t)dt}{a(t)},\label{displ}
\end{equation}
where the peculiar velocity is $v(t)=v_ia(t_i)/a(t)$. We require
that the displacement $\Delta r$ is smaller in comparison with the loop's radius
$l/(2\pi)$. From this condition and (\ref{displ}) we obtain the restriction
on the loop's velocity
\begin{equation}
v_it_i\ln(t_d/t_i)<l/(2\pi).\label{cond}
\end{equation}
For the probable parameters of the strings
$t_d/t_i\simeq2\times10^5$ and the dependence in (\ref{cond}) is
only logarithmic. Therefore, by using (\ref{distrv}) we can
estimate the probability of the low velocity loop formation as
\begin{equation}
P_{\rm lv}\sim\frac{(2/\pi)^{1/2}v_i^3}{3\langle
v_i^2\rangle^{3/2}}\simeq2\times10^{-7}.\label{problowv}
\end{equation}
As we will show below even the tiny fraction (\ref{problowv}) of the
formed loops may produce superdense clumps and observable
annihilation signals.

Now we consider the rocket effect. Velocity of the loop grows
linearly with time $v_r=3\Gamma_PG\mu t/(5l)$,
where $\Gamma_P\sim10$ \cite{VilSheBbook}. Turnaround moment of the clump corresponds to
$t_{TA}^2\simeq500t_i^2$, and the relative displacement of the loop during clump formation is
\begin{equation}
\frac{1}{l}\int\limits_{t_i}^{t_{TA}}v_rdt\simeq1.5\times10^{-4}\mu_{-8}\alpha_{0.1}^{-1}\ll1,
\end{equation}
where $\mu_{-8}\equiv G\mu/(10^{-8}c^2)$. Therefore, for the small loops formed at radiation era the large rocket displacements are not achieved.

\section{Evolution of clumps around evaporating loops}
\label{evolsec}

We solve the same equation as eq.~(2.7) in \cite{KolTka94}
\begin{equation}
x(x+1)\frac{d^2b}{dx^2}+ \left[1+\frac{3}{2}x\right]
\frac{db}{dx}+\frac{1}{2} \left[ \frac{1+\Phi}{b^2}-b
\right]=0,\label{bigeq}
\end{equation}
where $x=a(t)/a_{\mathrm{eq}}$ is used as independent variable, $r=ab(x)\xi$ is the physical radius of the spherical shell and $\xi$ is its comoving coordinate, $\Phi$ is the density
perturbation of DM $\delta\rho/{\bar\rho}$ inside the spherical shell. This equation describes the evolution of the clump's radius $r$ in terms of the function $b(x)$. The only quantity
one need to modify is the $\Phi$. In difference with \cite{KolTka94} we
allow the dependence of $\Phi$ on the time: steady decrease in the
continuous evaporation approximation and step-like in the fast decay
approximation. The evolution of clump stops when $dr/dt=0$ or
equivalently $db/dx=-b/x$ \cite{KolTka94}. The density and the radius
of the clump at the moment of its maximum expansion are
\begin{equation}
 \rho_{\rm max}=\rho_{\rm eq}x_{\rm max}^{-3}b_{\rm max}^{-3}, \quad
 R_{\rm max}=\left(\frac{3M}{4\pi\rho_{\rm max}}
\right)^{1/3},
 \label{rmax}
\end{equation}
where $b_{\rm max}$ and $x_{\rm max}$ are the values at the moment
of the stop.  After the turnaround the clump virializes by contracting twice in radius,
and the resulting density increases by factor 8 in comparison with (\ref{rmax}).

Let us name the spherical region with a volume
$(4\pi/3)\cdot(l/2\pi)^3$ as a ``string volume''. Then the fraction of
string mass $M_l=\mu l$ to the mass $M_{\rm DM}^l$ of DM inside
the string volume at the moment of the string birth $t_i=l/(\alpha c)$ is simply
\begin{equation}
\left. \frac{M_l}{M_{\rm DM}^l} \right|_{t=t_i}=\left(
\frac{M_l}{M_{\beta}}\right)^{-1/2}, \label{fracmsmdm}
\end{equation}
where $M_{\beta}=1.6\times10^3\mu_{-8}^3\alpha_{0.1}^{-3}M_{\odot}$. The
(\ref{fracmsmdm}) was calculated from the law of DM density evolution $\propto a^{-3}$
according to the known solutions of the Friedmann equations. The
fraction (\ref{fracmsmdm}) gives also the value $\Phi$ of the density
perturbation inside the string volume at the moment of string
birth $t_i$, and in the most interesting cases $\Phi\gg1$, with the production of the superdense clumps. The strings with $M_l=M_{\rm DM}^l(t=t_i)$ born at the time
$t_{\beta}=3.9\times10^{-6}\mu_{-8}^2\alpha_{0.1}^{-4}t_{\rm eq}$
($x_{\beta}=2\times10^{-3}\mu_{-8}\alpha_{0.1}^{-2}$). We consider
only the most dense central part of the clump inside the string
volume, where the annihilation proceeds most effectively. This central
region of the clump can be refereed as a clump core. The oscillation of string itself does not
distort the core because the linear segment of the string has small gravitational potential. The only
common (mean over the string volume) potential of the loop attracts the DM. The outer
regions of the clump form through the secondary accretion of DM
and have the density profile $\rho(r)\propto r^{-9/4}$ at the
sufficiently high distance from the center of the clump. Therefore
the annihilation concentrates near the clump core.

\section{Continuous evaporation and fast decay approximations}

The characteristic loop lifetime due to gravitational waves
emission is $\tau\simeq lc/(G\mu\Gamma)$, where $\Gamma\sim50$ is
a numerical coefficient \cite{Gambig}. One
possibility is that the loop losses mass continuously
according to mean equation $dM_l/dt=-\Gamma G\mu^2/c$, but the
more reliable approach is to assume the sudden decay after
the time interval $\tau$ from the birth moment $t_i$. At $t>t_i+\tau$ the loop's
configuration substantially changed, so the loop is likely to
self intersect. The resulting daughter loops will fly away at high
speeds. We use the last approximation henceforth as the main but also present the results
for the continuous evaporation approximation for comparison.

DM clumps formation at the RD stage was explored in
\cite{KolTka94}. In the particular case for the clumps which are seeded
by loops of cosmic strings one have $\Phi\simeq M_l/M$, but in many cases
the maximum density is only $\rho_{\rm
cl}\simeq140\rho_{\rm eq}$ due to adiabatic expansion of already
formed clump during the seed loop gravitational evaporation \cite{KolTka94}.
Really, the universal Poincare adiabatic invariant conservation
$J=\oint\sum p_idq_i=const$ for the clump with additional loop mass inside implies
$M_{\rm tot}R=const$ or $\rho_{\rm cl}\propto M_{\rm tot}^{-3}$, where
$M_{\rm tot}=M_l(t)+M_{\rm DM}$ is the total mass of the loop and DM.
For the constant mass loop the clump forms with the density $\rho_{\rm
cl}\simeq140\rho_{\rm eq}(M_l/M_{\rm DM})^3$ (see eq.~(3.4) in \cite{KolTka94})).
After the subsequent loop decay the density lowered due to adiabatic invariant
in proportion $\simeq(M_{\rm DM}/M_l)^3$ till the value $\rho_{\rm
cl}\simeq140\rho_{\rm eq}$.

We argue that the above mentioned argument of the adiabatic invariant conservation
is not applicable, if the loop decay occurs before turnaround moment (detachment from
the cosmological expansion and the clump virialization). Really, in this
case DM particles move not at the orbits around the loop but along the
radial trajectories. The loop's decay leads only to the change of
the particles acceleration. The evolution of clump slows down but
continues under the influence of velocities $db/dt$ obtained before
the decay. We can estimate the processes by the following manner.
For the clumps under consideration the conditions $x\ll1$,
$\Phi\gg1$ and $\Phi x\ll1$ are valid almost all the time before
turnaround. Initially the evolution goes due to large value of
$\Phi$ and the initial velocities $db/dx$ at $t=t_i$ are not important.
At this stage one can neglect the first term in (\ref{bigeq}) and
the approximate solution is $b\simeq1-x\Phi/2$ \cite{KolTka94}. If
the turnaround occurs before $t_d$ the moment of turnaround can be
estimated as $x_{\rm TA}\sim1/\Phi$ \cite{KolTka94}. In the
opposite case $x_d<x_{\rm TA}$ just after loop decay (at $x=x_d$)
we must put $\Phi=0$ in (\ref{bigeq}) and the velocity at this
moment $db/dx=-\Phi/2$ becomes greater then the last term in
(\ref{bigeq}). At $x>x_d$ we can neglect the last term but leave
the first one. This leads to the red-shifting of velocity as
\begin{equation}
\frac{db}{dx}=-\frac{\Phi x_d}{2x}
\end{equation}
and to the corresponding logarithmic decrease of $b$. From the condition
$db/dx=-b/x$ we obtain the new turnaround moment:
\begin{equation}
x_{\rm TA}\sim x_d\exp\left(\frac{2(1-\Phi x_d)}{\Phi
x_d}\right).
 \label{xtanew}
\end{equation}
We see that at the sufficiently small $\Phi x_d\ll1$ the clump may not
forms at all, because the $x_{\rm TA}$ will be exponentially large.
For the moderately small values $\Phi x_d\ll1$ the clump forms but
with small density. The value $\Phi x_d$ can be expressed as
\begin{equation}
\Phi
x_d\simeq0.9\mu_{-8}^{1/2}\alpha_{0.1}^{-3/2}\Gamma_{50}^{-1/2}.
\label{phixd}
\end{equation}
For the constant mass loop $x_{\rm TA}\sim1/\Phi$ \cite{KolTka94}
and (\ref{phixd}) is $\sim x_d/x_{\rm TA}$. This value is close to
unity at $\mu_{-8}\sim1$, so we expect the change in the character
of the clump formation process near $\mu_{-8}\sim1$.

In the continuous evaporation approximation the rate of loop mass
evaporation due to gravitational radiation is
$dM_l/dt=-\Gamma G\mu^2/c$.
After integration we have
\begin{equation}
M_l(t)=M_l(t_i)\left(
1-5\times10^{-6}\frac{\mu_{-8}\Gamma_{50}}{\alpha_{0.1}}\left[\frac{t}{t_i}-1\right]
\right),\label{mlevol}
\end{equation}
where the $M_l(t_i)=\mu\alpha t_i$ is the initial string mass at
the moment of its birth $t_i$.
One need to generalize the evolving seed mass as the
fuse of fluctuation growth. The only quantity one need to change
is $\Phi$. From the known solutions of the Fridmann equations
one has the dependence
$t=\tilde t(x)$ and by using (\ref{fracmsmdm}) we finally obtain
\begin{equation}
 \Phi(x,x_i)=\frac{2\times10^{-3}\mu_{-8}\alpha_{0.1}^{-2}}{x_i}
 \frac{M_l(\tilde t(x))}{M_l(\tilde t(x_i))}.
\end{equation}
This expression is valid for $\Phi\ge0$. If formally $\Phi<0$ we put
$\Phi=0$ in (\ref{bigeq}). This means that the string had totally
evaporated (its mass is zero) and the subsequent clump evolution
proceeds only under DM self gravity and due to inward velocity
boost which appeared before the string decay. The velocity boost
leads to the perturbation growth even after full evaporation of
the seed loop. The clump virializes with some density $\rho_{\rm cl}(t_{\rm TA})$.
If the string mass goes to zero after the turnaround the adiabatic expansion
of the clump occurs only due to the loop mass remnant $M_l(t_{\rm TA})$, and
the resulting clump density is
\begin{equation}
\rho_{\rm cl}=\rho_{\rm cl}(t_{\rm TA})\left(\frac{M_{\rm DM}}{M_l(t_{\rm TA})+M_{\rm DM}}\right)^3=
\frac{\rho_{\rm cl}(t_{\rm TA})}{(1+\Phi(x_{\rm TA}))^3}.
\label{oneplusp}
\end{equation}
In the case $t_d>t_{\rm TA}$ this density is greater in comparison with density $\rho_{\rm
cl}\simeq140\rho_{\rm eq}$ in the fast decay approximation.

\section{Numerical results}
\label{subnumres}

\begin{figure}[t]
\begin{center}
\includegraphics[angle=0,width=0.9\textwidth]{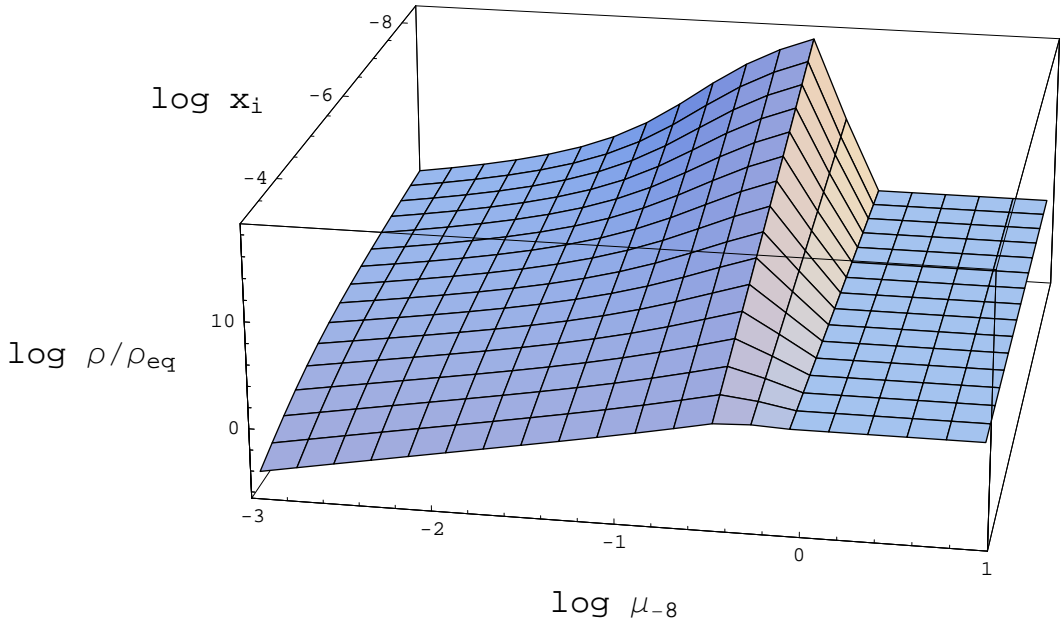}
\end{center}
\caption{Clump density $\rho$ in the units of density at
matter-radiation equality $\rho_{\rm eq}$ in dependence on the
loop birth moment $x_i$ and parameter
$\mu_{-8}=G\mu/(10^{-8}c^2)$. The break of the surface down to
value $\rho=140\rho_{\rm eq}$ corresponds to the proximity of
turnaround and loop decay moments.} \label{fig3dl}
\end{figure}

We solve Eq.~(\ref{bigeq}) numerically in the two above approximations. In the approximation of
fast decay the loop decay at the moment $t_d=t_i+\tau$. This means that at
$t<t_d$ the mass of the string is constant, but at $t>t_d$ the
string had totally disappeared (its mass is zero), and the
subsequent clump evolution proceeds only under DM self gravity and
due to inward velocity boost which appeared before the string
decay. The velocity boost leads to the perturbation growth even
after full evaporation of the seed loop.

We consider only the most dense central region of the clumps, which
gives the main contribution to the annihilation signal. These are regions
inside the string volumes. As the first approximation we consider
these regions as homogeneous. We find the density of the clump in
dependence of $x_i$ and $\mu$. The turnaround moment is calculated
numerically from the condition $db/dx=-b/x$ and the solution of
(\ref{bigeq}). Clumps density is obtained according to
(\ref{rmax}) and (\ref{oneplusp}). If the turnaround moment precedes the loop decay, we
put the resulting clump density $\rho=140\rho_{\rm eq}$ according
to adiabatic invariant argument conservation of \cite{KolTka94}.

The results of calculations for clumps density in the fast decay approximation are shown at
the figure~\ref{fig3dl}. As it was expected from (\ref{phixd}), the
condition $x_{\rm TA}\simeq x_d$ is satisfied near $\mu_{-8}\sim1$,
and the regime of clump formation changes near $\mu_{-8}\sim1$
because at larger $\mu_{-8}$ the turnaround occurs before the loop
decay. The similar figure can be presented for the continuous evaporation approximation,
but with smoother surface break and with greater density of the clumps.

\section{Loops and clumps distributions}
\label{distrsec}

The length distribution of cosmic strings' loops in the interconnecting network 
was obtained in \cite{OluVil06} in the form
\begin{equation}
dn_{\rm loop}=\frac{Ndl}{c^{3/2}t^{3/2}l^{5/2}}, \label{loopsdis}
\end{equation}
where $N\sim2$. The evaporating mass cutoff must be superimposed
on the distribution (\ref{loopsdis}) at the every particular time. If
we neglect (temporary) the loop evaporation, then the mass
fraction of the universe in the form of loops at the time $t_{\rm eq}$ is
\begin{equation}
\frac{d \rho_l(t_{\rm eq})}{\rho_{\rm
eq}}=0.042\mu_{-8}^{3/2}\left(\frac{M_l}{M_{\odot}}\right)^{-3/2}\frac{dM_l}{M_{\odot}}.
\label{ld2}
\end{equation}
In terms of cosmological density of clumps (a fraction of DM mass in
the form of clumps) the distribution (\ref{ld2}) (by using
(\ref{fracmsmdm})) can be rewritten as:
\begin{equation}
d\xi_{\rm cl}\simeq \frac {d\rho_l(t_{\rm eq})} {\rho_{\rm eq}}
\left(\frac{M_l}{M_{\beta}}\right)^{1/2}P_{\rm lv},\label{ld3}
\end{equation}
where $P_{\rm lv}$ is given by (\ref{problowv}). Strings decay but
the clumps survive, therefore there is no need to cut of the
clumps mass spectrum at the string evaporation scale, and the
(\ref{ld3}) is the real distribution of clumps at MD epoch. The
density of these clumps was calculated in the
Section~\ref{subnumres}.

The low mass cut of of the clumps distribution is determined by the process of
kinetic decoupling of the DM particles. At earlier times the DM particles strongly frozen in the radiation and do not move toward the loop. In contrast to the ordinary inflationary density perturbations the diffusion and free streaming effects are not important for the minimum mass of the clumps. This is because the forming clump subjected mainly by the strong gravitational
pull of the central loop and evolve nonlinearly long before the equality moment $t_{\rm eq}$. The kinetic decoupling temperature for ordinary neutralino weakly depends on the particle mass $T_d\propto m_{\chi}^{1/4}$ and, for example, for $m_{100}\equiv m_\chi/(100\mbox{~GeV})=1$ and for typical SUSY parameters $T_d\simeq25$~GeV with corresponding cosmological time $t_d\simeq1.2\times10^{-3}$~s. The loops which formed at the moment $t_d$ have masses $M_{l,{\rm min}}=2.5\times10^{-7}m_{100}^{-1/2}\mu_{-8}\alpha_{0.1}M_{\odot}$ and the minimum clump's mass is therefore $M_{\rm cl,min}=M_{l,{\rm min}}^{3/2}/M_{\beta}^{1/2} \simeq2\times10^{-15}m_{100}^{-3/4}\alpha_{0.1}^3M_{\odot}$ according to (\ref{fracmsmdm}).  These minimum mass clumps can reach densities $\rho_{\rm cl}\sim3\times10^{-4}$~g~cm$^{-3}$ if $\mu_{-8}\simeq0.4$ (see figure~\ref{fig3dl}).

\section{DM annihilation}
\label{annsec}

\begin{figure}[t]
\begin{center}
\includegraphics[angle=0,width=0.9\textwidth]{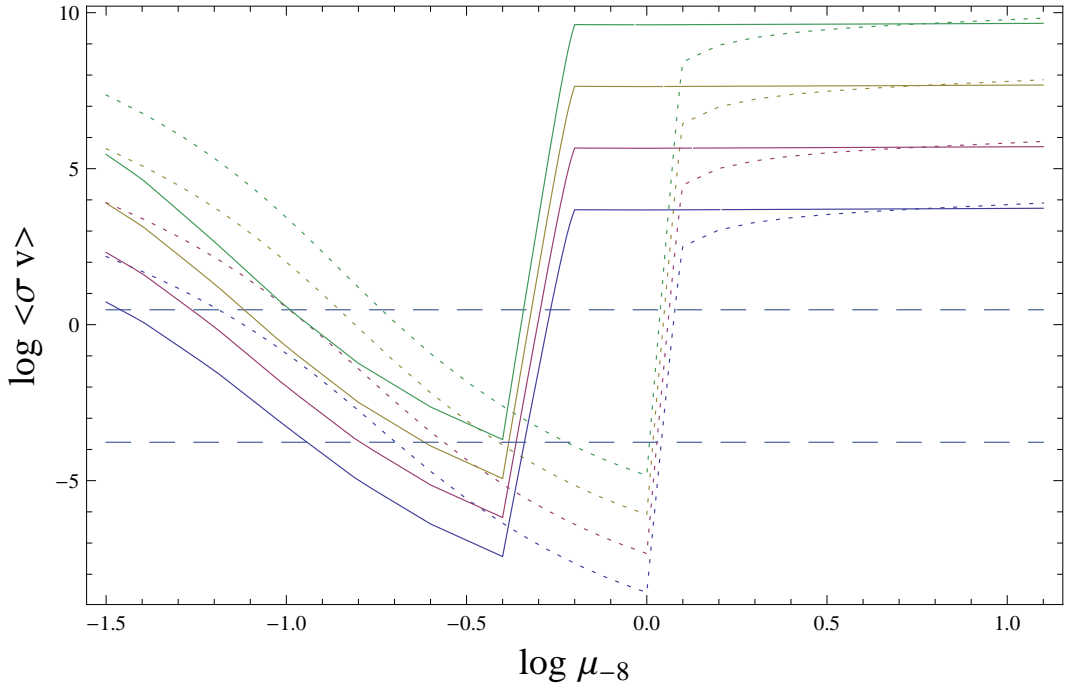}
\end{center}
\caption{Upper limits on $\langle\sigma v\rangle$ (in units $10^{-26}$~cm$^3$~s$^{-1}$) in
dependence of the string parameter $\mu_{-8}=G\mu/(10^{-8}c^2)$. The solid lines show the limits for the masses of DM particles (from up to down) $m_{\chi}=10$~TeV, $1$~TeV, $100$~GeV and $10$~GeV in the fast decay approximation. The limits
were obtained from the comparison of the calculated signals and the Fermi-LAT data.
The upper and lower horizontal dashed lines show the typical and minimal possible cross-section values, respectively. The dotted lines show the upper limits in the continuous evaporation approximation for the same masses.}
\label{figannl}
\end{figure}

The clumps under consideration have very large densities and the gamma-ray flux from DM annihilation inside the clumps may exceed the observational limits for some values of string parameter $\mu_{-8}$. Let us consider the neutralino
(the most popular DM candidate) annihilation in the clumps. Annihilation rate of neutralino in a single clump $\dot N_{\rm ann}=2\eta_{\pi^0}4\pi\langle\sigma
v\rangle\int\limits_{0}^{R}n_{\chi}^2r^2dr$,
where $\eta_{\pi^0}\sim10$ is the neutral pion multiplicity, $n_{\chi}$ is the number density of
particles inside clump, $R\simeq(3M/4\pi\rho)^{1/3}$ and $\langle\sigma
v\rangle$ is the annihilation cross-section (averaged product with velocity). We consider the annihilation channel with $\pi^0$ productions and decays $\pi^0\to2\gamma$.
The cumulative gamma-ray signal from the clumps in the angular direction $\psi$ with respect to Galactic center can be expressed as
\begin{equation}
J_{\gamma}(E>m_{\pi^0}/2,\psi)=1.9\times10^{-10}\left(\frac{m_{\chi}}{100
\mbox{~GeV}}\right)^{-2}\frac{\langle\sigma v\rangle}{10^{-26}\mbox{~cm$^3$s$^{-1}$}}
\langle J(\psi)\rangle_{\Delta\Omega},
\label{igam}
\end{equation}
where
\begin{equation}
\langle J(\psi)\rangle_{\Delta\Omega}=\int d\xi_{\rm cl}\left(\frac{\rho_{cl}}{0.3\mbox{~GeV~cm$^{-3}$}}\right)
\int\limits_{l.o.s.}\frac{dL}{8.5\mbox{~kpc}}
\left(\frac{\rho_H(r)}{0.3\mbox{~GeV~cm$^{-3}$}}\right),
\label{jpsi}
\end{equation}
and the last integration goes along the line of sight.
For the halo density profile $\rho_{\rm H}(r)$ we use the NFW profile
\cite{NFW} with the scale $a=20$~kpc, halo mass $M_h=10^{12}M_{\odot}$
and virial radius $R_h=200$~kpc. The lower mass limits in the integration $d\xi_{\rm cl}$ was estimated in the previous section. This limit weakly depends on $m_{\chi}$ through the $T_d(m_{\chi})$ dependence. We took $M_{l,\rm max} \simeq1.6\times10^3\mu_{-8}^3M_{\odot}$ (this seed mass corresponds to the clump's formation time near $t_{\rm eq}$) as the upper limits of the integration, and the dependence of the final result on $M_{l,\rm max}$ is weak.

We compare the calculated signals with Fermi-LAT diffuse extragalactic gamma-ray background
$J_{\rm obs}(E>m_{\pi^0}/2)=1.8\times10^{-5}$~cm$^{-2}$~s$^{-1}$~sr$^{-1}$ \cite{Fermi3603}.
To obtain the most conservative limit we compare $J_{\rm obs}$ with the calculated signal in the anti-center direction $\psi=\pi$. It gives the upper limit on $\langle\sigma v\rangle$ in dependence of $\mu_{-8}=G\mu/(10^{-8}c^2)$. The results are shown at figure~\ref{figannl}. We consider the several values of the neutralino mass: $m_{\chi}=10$~TeV, $1$~TeV, $100$~GeV and $10$~GeV. The mass $m_{\chi}$ influences the result mainly through the factor $m_{\chi}^{-2}$ under the integral (\ref{igam}) and through the low mass limit in the loops distribution, which weakly depends om $m_{\chi}$.

For example, for the mass $m_{\chi}=100$~GeV in the case of typical neutralino cross-section $\langle\sigma v\rangle\simeq3\times10^{-26}~\mbox{cm}^3\mbox{s}^{-1}$ (this value corresponds to the thermal production of DM particles) the limit excludes the range of parameters
$0.05<\mu_{-8}<0.51$ in fast decay approximation and $0.1<\mu_{-8}<1.16$ in the continuous evaporation approximation. If we take the minimal allowed value
$\langle\sigma v\rangle=1.7\times 10^{-30}m_{100}^{-2}~\mbox{cm}^3\mbox{s}^{-1}$ \cite{BerBotMig96} and $m_{\chi}=100$~GeV the excluded regions are $0.16<\mu_{-8}<0.43$ and $0.27<\mu_{-8}<1.07$ in the same two approximations.

\section{Conclusions}
\label{discussionsec}

In this work we explore the possibility of dark matter annihilation in
the dense cores of clumps that were seeded by cosmic string loops before the matter-radiation equality. We calculate the evolution of the clumps around evaporating loops. Only the
low-velocity loops result in the clumps formation. At the same time, even the low-velocity tail of the loop distribution produces DM clumps with the observable signature in the annihilation products.
The adiabatic argument conservation doesn't prevent the formation of clumps with densities $\rho_{\rm cl}\gg140\rho_{\rm eq}$, if the decay of the loops occurs before the time of the clumps virialization. Therefore the DM clumps produced by the loops can be the very dense objects with high cumulative luminosity in the gamma-rays.

The combined constraints on the loops parameters ($\mu$ and the distribution over lengths) and parameters of DM particles were obtained. For the 100~GeV neutralino DM the range $5\times10^{-10}<G\mu/c^2<5.1\times10^{-9}$ was excluded because of the huge gamma-ray annihilation signal above the Fermi-LAT data for these parameters. Along with the preferable neutralino dark matter candidate with the mass 100~GeV, we consider the masses $10$~TeV, $1$~TeV, and $10$~GeV and explore the two limiting assumptions about the loop evolution: fast decay and continuous evaporation approximations. The corresponding constraints on the annihilational cross-section are shown at the figure~\ref{figannl}. We restrict ourself only by the preferable value $\alpha=0.1$. The results will change for different $\alpha$.

In the case of the cosmic superstrings, the reconnection probability can be $\ll 1$,
resulting in the much higher number density of loops. This could
lead to even stronger constraints in comparison with the presented in this paper.

The Fermi-LAT data are used as the upper limit. In principle the annihilation of DM in the clumps
can explain the observed signal for the particular values, for example $G\mu/c^2\simeq5\times10^{-10}$. The necessity of the DM annihilation can arise if the ordinary astrophysical sources give too small signal in comparison with the observations.

\acknowledgments

We thank A. Vilenkin for the very useful suggestions and discussion.
This work was supported by the grants of the Russian Leading scientific schools 3517.2010.2 and
Russian Foundation of the Basic Research 10-02-00635.

\end{document}